\documentclass[prl,twocolumn,showpacs,floatfixlongbibliography]{revtex4-2}
\usepackage{epsfig}
\usepackage{amsmath}  
\usepackage{graphicx}
\usepackage{epsfig,latexsym,array}

\newcommand{\be}{\begin{equation}}
\newcommand{\ee}{\end{equation}}
\newcommand{\ba}{\begin{eqnarray}}
\newcommand{\ea}{\end{eqnarray}}
\newcommand{\benn}{\begin{displaymath}}
\newcommand{\eenn}{\end{displaymath}}
\renewcommand{\d}[1]{{\rm d}#1}

\renewcommand{\vec}[1]{\mbox{\boldmath $#1$}}

\newcommand{\JA}{Josephson-Anderson relation}
\reversemarginpar
\newcommand{\nonote}[1]{ } 

\newcommand{\deleted}[1]{\color{red} #1 \color{black}}

\renewcommand{\deleted}[1]{}
\begin{document} 
\title{First observation of the Josephson-Anderson relation in experiments on hydrodynamic drag
}
\author{Nicola Savelli, Ali R Khojasteh, {Abel-John} Buchner, Jerry
Westerweel and Willem van de Water}

\affiliation{
Laboratory for Aero- and Hydrodynamics, Delft University of Technology
and J.M. Burgers Centre for Fluid Dynamics, 2628 CD Delft, The
Netherlands}
\email[Corresponding author: ]{w.vandewater@tudelft.nl}
%
\date{31 March 2025}
\begin{abstract}
We verify a recent prediction (Eq. 3.50 in G. L. Eyink, Phys. Rev. X
{\bf 11}, 031054 (2021)) for the drag on an object moving through a
fluid.  In this prediction the velocity field is decomposed into a
nonvortical (potential) and vortical contribution, and so is the
associated drag force.  
In the Josephson-Anderson relation the vortical contribution of the
drag force follows from the flux of vorticity traversing the
streamlines of the corresponding potential flow. The potential
component is directly determined by the plate acceleration and its
added mass.  The \JA\ is derived from the quantum description of
superfluids, but remarkably applies to the classical fluid in our
experiment.
In our experiment a flat plate is accelerated through water using a
robotic arm. This geometry is simple enough to allow analytic
potential flow streamlines.  The monitored plate position shows an
oscillatory component of the acceleration, which adds an additional
test of the \JA.
The instantaneous velocity field is measured using particle image
velocimetry.  It enables us to evaluate Eq.~3.50 from \cite{Eyink2021}
and compare its prediction to the measured drag force.  
We find excellent agreement, and, most remarkably find that the added
mass contribution to the drag force still stands out after the
flow has turned vortical.
We finally comment on the requirements on the experimental techniques
for evaluating the \JA. 
\end{abstract}
\maketitle
%

{\it Introduction.---}
In a recent paper \citet{Eyink2021} explains the drag on an object
moving through a fluid by the flux of vorticity across streamlines of
the corresponding potential flow \citep{Eyink2024,Kumar2024}.
The equation relating drag to vorticity, the \JA, derives from the
quantum description of superfluids, but remarkably applies to
classical fluids.  It provides a quantitative exact and instantaneous
link between drag force and the velocity field.
In the present context drag would be absent if there is no flux of
vorticity across streamlines. It is an interesting corollary to
d'Alembert's paradox in which a moving object experiences no drag in
stationary potential flow.  
%

\begin{figure*}[t]
\centering
\includegraphics[scale = 0.8]{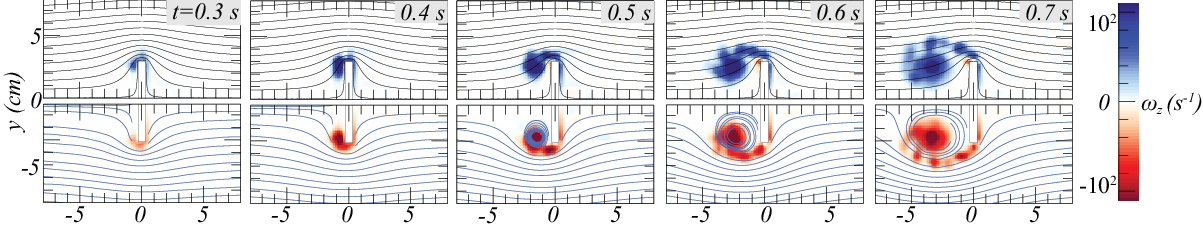}
\caption{
Evolution of the vorticity during acceleration with maximum ($a_{\rm
max} = 0.8\:{\rm m}{\rm s}^{-2}$) of the plate.  The final velocity ($U
= 0.3\:{\rm m}{\rm s}^{-1}$) is reached at $t = 0.6\:{\rm s}$.  
The upper half of all panels shows the streamlines of the potential flow, which are
symetric in $y$.  The lower half shows the true experimental streamlines (in blue).
The essence of the \JA\ is that drag ensues when vorticity, carried
by the true flow, crosses the streamlines of the potential flow.
Animation is in the Supplemental Material \cite{supplement}. 
%
}
\label{fig.strip}
\end{figure*}

The idea that the drag force on an object is related to the fate of
vorticity as it is ``washed away'' from the object, may be traced
back to J.M. Burgers \citep{Biesheuvel2006,Burgers1920}.  It has
spawned recent research on the connection between vorticity dynamics
and forces, for example to quantify and manipulate forces and
associate force events to vorticity events
\citep{Graham2019,Corkery2019,Gehlert2023}. Establishing this
connection became possible using recent advances in precisely mapping
the velocity field using Particle Image Velocimetry (PIV) and
robotized object manipulation \citep{EJ2019,Jesse2023}.

In our experiment we drag a plate through still fluid, vary its
acceleration, measure the drag force and map the velocity field. The
question is whether the \JA\ can predict the drag force in an
experiment using a measured velocity field.  In contrast to other
methods \cite{Rival2017} to infer forces from measured velocity
fields, neither pressure nor time information is needed: the force is
computed from snapshots of the velocity field.

A series of snapshots of the vorticity in our experiment is shown in
Fig.\ \ref{fig.strip}.  At time $t \approx 0.5\:{\rm s}$ a vortex
pair detaches from the plate edges.  It appears that at this time the
drag force reaches a maximum.

%
%
The essence of the \JA\ is the separation of the velocity field $\vec{u}$
into a potential velocity $\vec{u}_\phi$ and a rotational component
$\vec{u}_\omega$, $\vec{u} = \vec{u}_\phi + \vec{u}_\omega$.  The
irrotational (potential) part $\vec{u}_\phi$ derives from a potential
$\phi$, $\vec{u} = \nabla \phi$ and satisfies the no-penetration
condition at the plate boundary, and in- and outflow conditions at
the boundaries of the experimental domain, which are assumed far away
from the plate.  It guarantees a unique solution of the Laplace
equation for $\phi(\vec{x})$. 

Although d'Alembert's paradox precludes drag in stationary potential
flow, an accelerated object experiences a drag force $F_\phi = -m_a
\d U / \d t$, with $U(t)$ the instantaneous velocity of the object, 
and $m_a$ the added mass.
The added mass represents the equivalent mass of the fluid that is set into
motion when the plate is accelerated from rest \citep{Corkery2019}.
Its context is potential flow; its value is determined by the shape
of the object and the direction of motion.  

When an object is accelerated from rest, and vorticity has not yet
transpired, $\vec{u} \approx \vec{u}_\phi$ at the initial stage of
the motion.
%
One would naively think that the separation of the velocity field 
$\vec{u} = \vec{u}_\phi + \vec{u}_\omega$ and the concept of added
mass no longer makes sense for developed flow.  
%
%
The \JA\ suggests that the decomposition of the velocity field 
$\vec{u} = \vec{u}_\phi + \vec{u}_\omega$, with the associated
decomposition of the drag force, $\vec{F} = \vec{F}_\phi +
\vec{F}_\omega$ remains relevant, irrespective of the flow
development away from potential flow.  Below we present striking
evidence for this.

\begin{figure}[t]
\centering
\includegraphics[scale = 0.8]{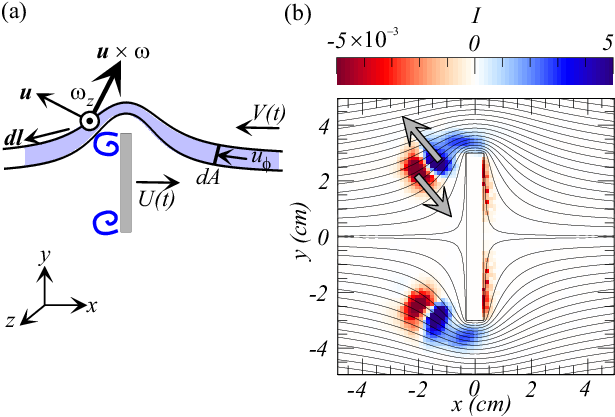}
\caption{
(a) Illustration of the JA relation as applied to our experiment in
which a plate moving to the right with velocity $U(t)$ sheds
vortices. The velocity $V(t)$ in Eq.\ \ref{eq.ja1} is $V(t) = -U(t)$.
Shown is a streamline with tangent vector $\d \vec{l}$.  At the upper
half of the plate the vorticity is positive, then for velocities
directed outward through a streamline, $\vec{u} \times \vec{\omega}$
points upward against the direction of $\d \vec{l}$, with a positive
contribution to the drag force.  
The flux $u_\phi\: \d A$ of potential flow between streamlines is the
difference between the values of the associated streamfunctions.
(b) Integrand $I = \d J \: \left(\vec{u}\times\vec{\omega} - \nu
\nabla \times \vec{\omega}\right) \cdot \d \vec{l}$ in the JA
relation Eq.\ \ref{eq.ja1}.  Its sign is such that positive values
add to drag, whereas negative values diminish drag.  
There is no contribution to drag when $\vec{u}$ is directed along
a streamline.  
The two arrows indicate the flux of vorticity across streamlines.
Equation\ (\ref{eq.ja1}) is evaluated using four times as
many streamlines as shown in panel (b).
%
Animation is in the Supplemental Material \cite{supplement}.
}
\label{fig.integrand}
\label{fig.ja1}
\end{figure}

The \JA\ expresses the vortex drag $\vec{F}_\omega$ on an object in
terms of the flux of vorticity across streamlines of the potential
flow field $\vec{u}_\phi$.  It was derived from the change of
energies in the potential and rotational flow in a channel by
\citet{Huggins1970}, and rederived for a moving body by
\citet{Eyink2021}.  The associated transfer of energy between both
components is the work done by an object against drag forces.
%
%
In our case, we apply it to a plate that moves horizontally with velocity $U(t)$ through still water. We then consider the plate's reference frame, such that the plate appears to be stationary and submerged in a flow with
velocity $\vec{V}(t)$ far from the plate, so that
$V_x(t) = -U(t)$.
The contribution $\vec{F}_\omega$ to the drag force due to vorticity
is then \citep{Eyink2021}
\be
\vec{F}_\omega \cdot \vec{V}(t) = \int \d J \int 
\left(\vec{u}\times\vec{\omega} - \nu \nabla \times
\vec{\omega}\right) \cdot \d \vec{l}
\label{eq.ja1}
\ee
with $\d \vec{l}$ pointing along the streamline.  The inner
integration is along a streamline, while in the outer integration $\d
J$ is the mass flux
within streamtubes,
$\d J = \rho \: \vec{u}_\phi \cdot \d \vec{A}$,
so that Eq.\ (\ref{eq.ja1}) represents the flux of vorticity across
potential mass current \citep{Eyink2021,Eyink2024}.
In 2D flows, $\d J$ is the mass flux between two streamlines.
The \JA\ holds for flow that is described by the incompressible
Navier Stokes equation in a domain that extends to infinity.
The \JA\ Eq.\ \ref{eq.ja1} is illustrated in Fig.\ \ref{fig.ja1}(a)
for the term $\vec{u} \times \vec{\omega}$ in Eq.\ \ref{eq.ja1}.  The
integrand in Eq.\ (\ref{eq.ja1}), measured in our experiment is shown
in Fig.\ \ref{fig.ja1}(b).

The unique and beautiful equation \ref{eq.ja1} is {\em
instantaneous}, providing an instantaneous prediction of the drag
force $\vec{F}_\omega$ from a measured velocity field, unlike other
methods that involve time derivatives
\citep{Corkery2019,Gehlert2023}, or solution of the Poisson equation
\citep{Oudheusden2013}

The decomposition $\vec{u} = \vec{u}_\phi + \vec{u}_\omega$ was also
discussed by \citet{Li1986} in the context of drag in ocean waves
and by \citet{Graham2019} who stresses that the irrotational velocity
field $\vec{u}_\phi$ acts upon the vortical forces $\vec{F}_\omega$
and suggests to derive this interaction using the analysis in
\citep{Howe1995,Eldredge2010,Limacher2018}.

In this paper we compare the measured drag force on an accelerated
plate to the force $\vec{F} = \vec{F}_\phi + \vec{F}_\omega$, with
$\vec{F}_\phi$ the added mass force and $\vec{F}_\omega$ given by
Eq.\ (\ref{eq.ja1}).  The terms in Eq.\ \ref{eq.ja1} were computed
from the measured velocity field.
The potential flow streamlines for the rectangular cross section of
the plate and boundary condition $u_\phi = V(t)$ at the left edge of
the measurement domain are computed using the Schwarz-Christoffel
transform \citep{Driscoll2009}.  
The potential flow streamlines form the natural integration grid in
Eq.\ (\ref{eq.ja1}).  With 512 streamlines spanning the experimental
domain (streamline spacing $6.5\times 10^{-4}\:{\rm m}$ at $x =
-0.4\:{\rm m}$), the resulting integration grid is finer than the
discrete velocity grid from PIV (grid cell size $2.66\times 10^{-3}\:
{\rm m})$.  Varying the number of streamlines from 512 to 1024 does
not change the computed $\vec{F}_\omega$ significantly.

We analyze two experiments, where the plate is accelerated to the
same final velocity, $U = 0.3\:{\rm m}{\rm s}^{-1}$ with maximum
acceleration of $\d U / \d t = 0.8 \;\; \mbox{and}\;\; 1.6 \:{\rm
m}{\rm s}^{-2}$, respectively.  
For these two experiments we compare the measured hydrodynamic force
on the plate with the prediction from the \JA.
%

\vspace*{2ex}{\it Experiment.---}
%
%
We accelerate plates (size $l \times h \times w = 0.3 \times 0.06
\times 0.006 \;{\rm m}^3$) through initially quiescent water.  The
time-dependent velocity field is measured using particle image
velocimetry.  The measured velocity field is accurate enough to
compute the vorticity $\vec{\omega}(\vec{x}, t)$ and diffusion of
vorticity $-\nu \: \nabla \times \vec{\omega}(\vec{x}, t)$, with
$\nu$ the kinematic viscosity. At our Reynolds numbers ${\rm Re}
\approx 2 \times 10^4$, as based on the final velocity and the width
$h$ of the plate, the contribution of the diffusion term is
insignificant. 

\begin{figure}[t]
\centering
\includegraphics[scale = 0.8]{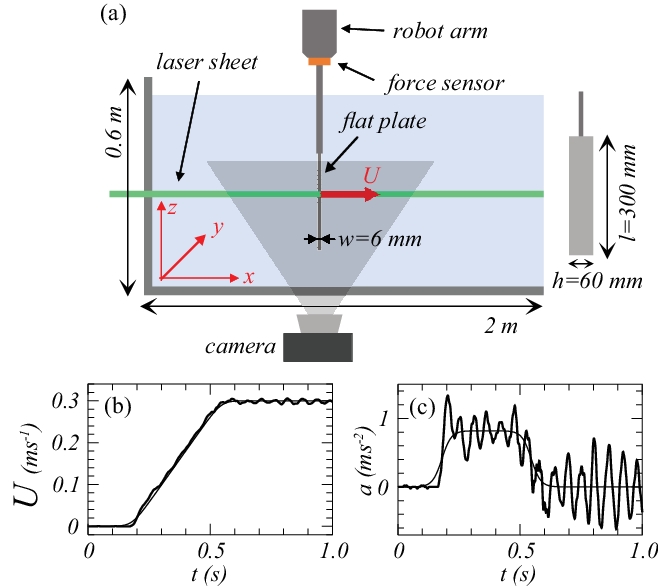}
\caption{
Experimental setup (not to scale).  (a) A rectangular plate is
accelerated through a large square tank ($2 \: {\rm m} \times 2 \:
{\rm m}$, depth $0.6\: {\rm m}$) filled with water.  A laser sheet
intersects the center of the plate, and illuminates tracer particles.
A camera (frame rate $1200 \:{\rm Hz}$) registers particle images.
Particle image velocimetry (PIV) is used to map the planar velocity
field.
(b,c) Time trace of the plate velocity and acceleration,
respectively. They were measured from optical tracking of the plate;
thin lines: nominal (as supplied to the robot).
}
\label{fig.setup}
\end{figure}

The setup is sketched in Fig.\ \ref{fig.setup}(a).  The aspect ratio
of the plate $l / h = 5$, and we measure the velocity field in a plane
bisecting the plate along its length $l$.  Hence, plate end effects
are sufficiently removed from the measurement plane, and the flow can
be considered as practically two-dimensional in the measurement plane
(i.e. $\partial w / \partial z \cong 0$).

The PIV system consists of a high-speed camera (Phantom Veo 640, 4
megapixels) with a frame rate of $1200\:{\rm Hz}$ and a laser (Litron
Nd:YLF, 150 W, thickness of laser sheet $3\:{\rm mm}$) that illuminates
fluorescent tracer particles (Cospheric, $65 \: \mu{\rm m}$ diameter)
in a $0.5\times0.5\: {\rm m}^2$ field of view. 
\nonote{We are working on an error estimate: either using DAVIS error,
and working it through Eq.1, or by estimating the variance of repeated
experiments. The uncertainty of the last section is unsatisfactory,
especially because we say that close streamlines do not contribute.}

The plate is moved by a gantry robot (Reis Robotics RL50); the drag force is registered by
a force/torque sensor (Schunk FTD-Gamma SI-32-2.5) with $6\times 10^{-3}\:{\rm N}$ force resolution.  It
is corrected for the mass of the plate and the drag of the supporting
strut.  The time dependent velocity of the plate in the first
experiment ($a = 0.8 \: {\rm m s}^{-2}$) is shown in Fig.\
\ref{fig.setup}(b).  The plate reaches its final velocity $U =
0.3\:{\rm m \: s}^{-1}$ at time $t = 0.6\: {\rm s}$.  In the frame of
reference considered here, the plate is stationary with its center at
$\vec{x} = 0$, while the fluid moves to the left with velocity $V(t)
= -U(t)$. 

%
%
%
%
Using sub-pixel interpolation and a pinhole camera model, the actual
plate position could be precisely traced from the PIV images, as the
plate is always in view.  The first derivative of the position trace
gives the plate velocity while the second derivative provides the
plate acceleration.  As Fig.\ \ref{fig.setup}(b,c) demonstrates, the
plate acceleration has an oscillatory component which could just be
discerned from the velocity signal.  Be it the consequence of the
finite stiffness of the plate mount, or the influence of the robot
(controller), we do not view it as an artifact, but as a fact.  It
allows us to study the response of the velocity and drag force to
(periodic) modulation.
The plate acceleration leads to the potential force contribution
${F}_{x, \phi} = m_a \d V / \d t$, with added mass $m_a$ for motion
in the $x-$direction, $m_a = \rho \; 3.35201 \ldots \; (h / 2)^2$
(Eq.\ (\ref{eq.am})).


\vspace*{2ex}{\it Results.---}
%
%
The steps leading to the force prediction are illustrated in Fig.\
\ref{fig.integrand} where we show snapshots of the integrand $I = \d
J \: \left(\vec{u}\times\vec{\omega} - \nu \nabla \times
\vec{\omega}\right) \cdot \d \vec{l}$ in Eq.\ \ref{eq.ja1}, and which
can be considered as a {\em heat map of drag}.  Notice that
$\omega_z$ is positive near the top of the plate, and negative at the
bottom, whereas the integrand in Eq.\ \ref{eq.ja1} switches sign when
$\vec{u}\times\vec{\omega}$ turns against the direction of the
potential flow velocity $\vec{u}_\phi$.
There is no contribution to drag when $\vec{u}$ is directed along
a streamline.  
The positive regions in Fig.\ \ref{fig.integrand}(b) correspond to
outward motion of vorticity across streamlines.

\begin{figure}[t]
\centering
\includegraphics[scale = 0.8]{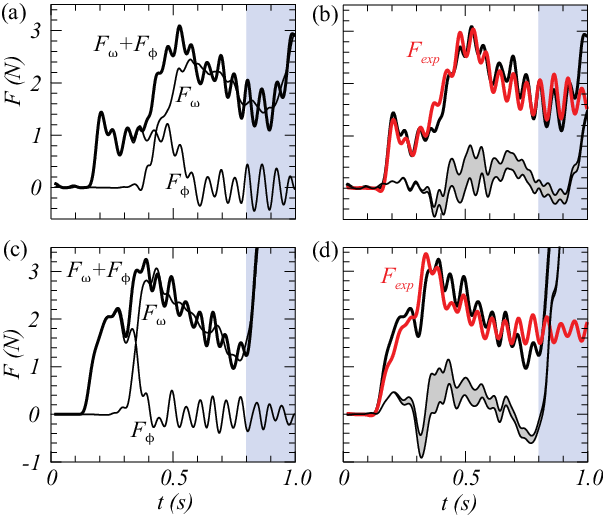}
\caption{
Measured (red lines in b, d) and computed hydrodynamic drag forces
for two acceleration scenarios: in panels (a,b) the maximum
acceleration is $a_{\rm max} = 0.8 \: {\rm m}{\rm s}^{-2}$, while in
(c,d) $a_{\rm max} = 1.6 \: {\rm m}{\rm s}^{-2}$.  Both scenarios end
with the same final velocity, $U = 0.3 \: {\rm m}{\rm s}^{-1}$.
%
%
(a,c) Drag forces computed from the measured velocity field using the
\JA\ Eq.\ref{eq.ja1}.  The two contributions: the potental force
$F_\phi$ and the vortical force $F_\omega$ as well as their sum are
indicated.
The computation of the drag forces uses the measured plate
accelerations (see Fig.\ \ref{fig.setup}(b,c)).
At long times, $t > 0.8 \:{\rm s}$ (shaded in the figure), the
flow turns three dimensional, and our approach, which is based on
planar flow, no longer applies. 
(b,d) Comparison of measured and computed forces.  Red lines:
$F_{\rm exp}$, black lines: $F_{\phi} + F_{\omega}$.  Gray lines:
difference $F_\phi + F_\omega - F_{\rm exp}$ between the computed
total force and experiment.  The width of the line represents an
estimate of the uncertainty obtained from multiple repetitions.
All results (time traces) have been filtered using a binomial filter
with root mean square width of $0.015 \: {\rm s}$.
}
\label{fig.forces}
\end{figure}

The the total drag force $\vec{F}_\omega + \vec{F}_\phi$
is compared to the measured force in Fig.\ \ref{fig.forces}.
The measured peak forces agree with the prediction model proposed by \citet{Jesse2023}.
A striking observation is that the sharp rise of the total drag force
at the start of the acceleration at $t = 0.2\:{\rm s}$ is completely
due to the potential force.  Until $t = 0.4\:{\rm s}$ and $t =
0.3\:{\rm s}$, for the cases $a_{\rm max} = 0.8 \:{\rm ms}^{-2}$ and
$a_{\rm max} = 1.6 \:{\rm ms}^{-2}$, respectively, the contribution
of the vortical force is negligible.  
The maximum of the total drag force is almost entirely due to the
vortical contribution $F_\omega$ and is reached {\em after} the added
mass force $F_\phi$ ceases. 

The measured force oscillations follow from the oscillating
acceleration and the added mass.  The excellent agreement even
extends to long times when the flow around the plate has turned
vortical and the concept of added mass, which is based on
irrotational flow, no longer applies strictly.


The modulatons of the acceleration could influence $\vec{u}_\phi$,
$\vec{u}_\omega$, or both, and could {\em via} them result in the
measured drag modulation.  As our results in Fig.\
\ref{fig.forces}(a, c) demonstrate, acceleration modulations are
mainly communicated through potential forces.
In Fig.\ \ref{fig.forces}(b, d) the computed and measured ($F_{\rm exp}$) drag forces
are compared.  The overall agreement between the measured force and
the prediction of the \JA\ is excellent,
but incidentally can be as large as 30\% for the experiment with the
largest acceleration.  The shown error estimate is discussed
below.


%
\nonote{Restitching and resampling the piv fields did not change the results.}

\vspace*{2ex}{\it Applying the \JA\ to experimental data.---}
Let us now discuss the possible errors incurred when applying Eq.\
(\ref{eq.ja1}) to our data.  
First, the \JA\ assumes an infinite domain, but the experiment domain
is neccessarily restricted.  The influence of the domain boundaries
can be learned in the experiment from evaluating Eq.\ (\ref{eq.ja1})
for the drag force on a smaller domain.  Clearly, the plate, being
the source of vorticity, should remain inside.  No significant
difference was found when reducing the linear dimensions by a factor
of 3.
%

Measurement of the velocity- and vorticity fields close to the plate
presents an experimental challenge. Tracer particles at both sides of
the plate are illuminated and used in the PIV analysis.
%
The finite resolution of PIV (velocity grid cell $2.66\times10^{-3}
\:{\rm m}$) affects the measurement of velocity gradients, which are
large close to the plate edges.  With a Reynolds number ${\rm Re} = h
\: U / \nu \approx 2\times10^4$, we estimate a boundary layer
thickness ${\rm Re}^{-1/2} \: h\approx 4\times 10^{-4} \:{\rm m}$;
which remains unresolved.
%
%
\nonote{Can we say more using Hiemenzs flow ?}

The relevance of vorticity close to the plate in computing the drag
force from Eq.\ (\ref{eq.ja1}) can be assessed by excluding vorticity
in a thin layer (width $5.2\:{\rm mm}$)
from the top and bottom edges (at $y = \pm h/2$) of the plate in the
integration in Eq.\ (\ref{eq.ja1}).  The comparison of the vortex
force computed from the full experimental domain to that from the
domain with this layer deleted leads to an estimate of the
uncertainty in the computed force.  This estimate is shown in Fig.\
\ref{fig.forces}(b,d).  Clearly, the estimated uncertainty is smaller
than the difference between computed and measured drag force,
pointing to other sources of error.

%
The potential flow streamlines form the grid used in the integration
in Eq.\ \ref{eq.ja1}: the ``backbone'' of the drag force.  The
streamlines close to the stagnation streamline (at $y = 0$) closely
hug the plate cross section and the integration over them depends on
the measured velocity (vorticity) field close to the plate.

\begin{figure}[t]
\centering
\includegraphics[scale = 0.8]{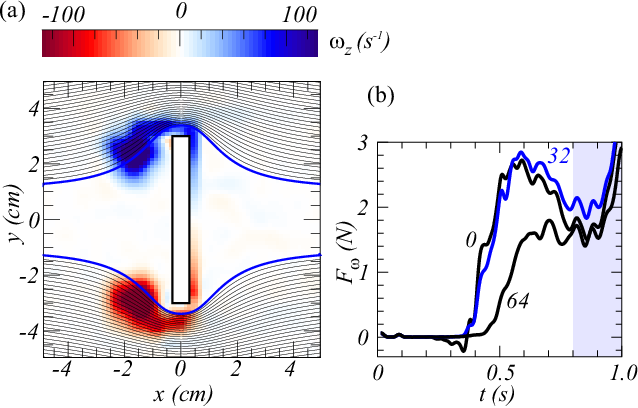}
\caption{
Dependendence of the computed vortex force $\vec{F}_\omega$ at $t =
0.5\: {\rm s}$ on the grid of streamlines, used as integration grid
in Eq.\ \ref{eq.ja1}. (a) Omitting the 32 streamlines (out of 512)
that are closest to the plate; color scale: vorticity $\omega_z$.
%
%
(b) Influence of the number of omitted streamlines on the vortical
force $\vec{F}_\omega$. For the curve indicated ``0'', all 512
streamlines were used in the integration; the case ``32'' is
illustrated in panel (a).
}
\label{fig.grace}
\end{figure}

The dependence of the computed force on these velocities can be
learned from omitting close streamlines from the integration in Eq.\
\ref{eq.ja1}. This is illustrated in Fig.\ \ref{fig.grace}(a),
while the resulting vortical force curve $\vec{F}_\omega$ is shown
in Fig.\ \ref{fig.grace}(b).  We conclude that the drag maximum is
caused by streamlines hugging the plate edges, demonstrating how the
\JA\ reveals the origin of drag.


Finally, our analysis is two-dimensional and is applied to a
two-dimensional cut of the velocity field.  
This is suggested by the aspect ratio $l / h = 5$ of our plate.  
At initial times a rectangular vortex ring is formed, of which we
observe a 2D cut.  
Two-dimensionality can be checked through the measured divergence of
the velocity field; it is progressively lost after $t \approx
0.8\:{\rm s}$.
A challenge remains the application of the \JA\ to a full 3D
experimental velocity field.


%
\vspace*{2ex}{\it Conclusion.---}
Both the measured and computed drag forces reach a maximum shortly,
when the acceleration starts to decline.  From our results it appears
that this maximum owes itself to the vortical force $\vec{F}_\omega$;
it is not an added mass (potential flow) effect. \citet{Jesse2023}
find that the force at this maximum scales with the square root of
the acceleration, which is explained by the diffusive growth of the
boundary layer on the plate.
%


A prime motivation of our work is to infer forces on an object from
an experimentally measured velocity field.  A great advantage of the
\JA\ is that the time dependence of the velocity field and the
pressure field are not required, however, velocity information close
to the object 
where the vorticity detaches
is needed.
%
%
\begin{acknowledgments}
{\it Acknowledgments --} The authors thank Gregory Eyink for helpful
comments and criticism.  
The expertise on PIV of Edwin Overmars was essential for this
project. 
This work is part of the ``ImpulsiveFlows'' project that has received
funding from the European Research Council (ERC) under the Horizon
2020 program (Grant No. 884778). A-J. Buchner was supported by a grant from the Netherlands Organisation for Scientific Research, NWO VENI 18176.
\end{acknowledgments}

\bibliography{jap}

\newpage
\centerline{\bf \large End Matter}
\appendix {\it Added mass--}
\label{sec:am}
The potential force contribution ${F}_{x, \phi} = m_a \d V / \d t$
needs the added mass $m_a$ for motion in the $x-$direction.  It was
computed for a plate with rectangular cross section $h \times w$ that
is accelerated through still fluid using the Schwarz-Christoffel
transform.  The added mass per unit length is
\be
   m_a = 
   \rho \int_S \phi\: {n}_x \: \d S =
   \rho \; 3.35201 \ldots \; (h / 2)^2,
\label{eq.am}   
\ee
with $\rho$ the density of the fluid, $\phi$ the unit potential field
(unit velocity at far away boundaries) and $\vec{n}$ the unit normal
on the plate surface.  It is very close to the added mass per unit
length $l$ of an infinitely thin plate (thickness $w = 0$) and width
$h$, which equals the added mass of a cylinder with radius $h / 2$,
\be
   m_a = 
   \rho \int_S \phi\: {n}_x \: \d S =
   \rho \; \pi \; (h / 2)^2 = \rho \; 3.14159 \ldots\; (h/2)^2.
\label{eq.am2}   
\ee
%
Still, it appears essential to use the streamlines of the thick plate
in the computation of Eq.\ (\ref{eq.ja1}) as usage of the thin plate
streamlines leads to a significantly smaller vortex force $F_\omega$.

\end{document}